# HOW TO MAKE THE DREAM COME TRUE: THE ASTRONOMERS' DATA MANIFESTO

*Ray P. Norris*
*CSIRO Australia Telescope, PO Box 76, Epping, NSW 1710, Australia*
*Email: Ray.Norris@csiro.au*

## ABSTRACT

*Astronomy is one of the most data-intensive of the sciences. Data technology is accelerating the quality and effectiveness of its research, and the rate of astronomical discovery is higher than ever. As a result, many view astronomy as being in a "Golden Age", and projects such as the Virtual Observatory are amongst the most ambitious data projects in any field of science. But these powerful tools will be impotent unless the data on which they operate are of matching quality. Astronomy, like other fields of science, therefore needs to establish and agree on a set of guiding principles for the management of astronomical data. To focus this process, we are constructing a "data manifesto", which proposes guidelines to maximise the rate and cost-effectiveness of scientific discovery.*

## 1. INTRODUCTION

The last few years have seen a revolution in the way astronomers use data. An astronomer can type the name of an object into a web page, and instantly view a wide range of observed data on that object, obtain references to all publications that mention it, and even produce plots of the spectral energy distribution (SED: Fig 1).

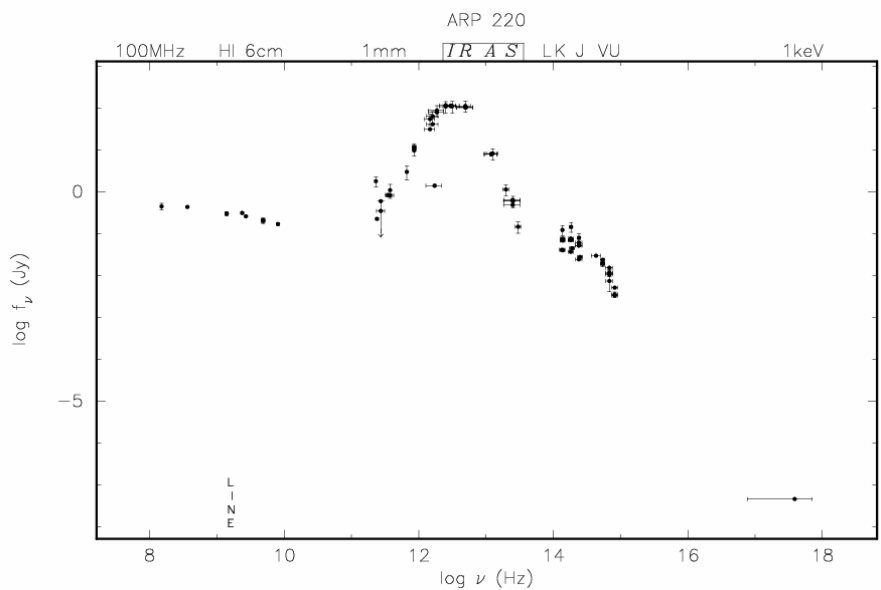

**Figure 1**: A typical Spectral Energy Distribution (SED) generated automatically by the NASA/IPAC Extragalactic Database (NED) using data collected by many different authors and instruments.

This SED includes data obtained from many different instruments using different technologies, calibration processes, and data formats, brought together in data centres that understand the instrument-specific metadata. All papers written in the astronomical journals are available on-line through a powerful engine that searches the entire body of astronomical literature. Many such papers will contain links to other publications and data.

The Virtual Observatory (VO) promises to place even more power at the hands of the astronomer, and with it the capability of accelerating the rate of scientific discovery. The VO will enable the astronomer to search all available databases in a region of sky, and superimpose or combine the images, or produce a plot comparing measurements on different instruments. Some of us dream even further. For example, I look forward to the day when I can move my mouse over an image I have just produced, and the VO dynamically gives me all available information, in the form of graphs, images, and literature, about the position underneath my cursor.

How do we turn these dreams and promises into reality? One requisite is obviously to build the necessary tools, services, and data structures, and the VO is doing just that. However, these tools will be ineffective without high-quality data on which to operate. While data from major international observatories, such as the European Southern Observatory and NASA's Great Observatories, are now freely available and managed in a way that is difficult to fault, much, perhaps most, of the remaining astronomical data and information are still relatively inaccessible. Even worse, some of these data are so poorly managed that they will be lost.

One of the reasons for poor data management is that many astronomers and observatory directors are unaware that good data management can generate good science, and that bad data management can inhibit the process of scientific discovery. Furthermore, in many areas there is not even a consensus on what constitutes good data management (Norris, 2005; Norris et al. 2006).

In an attempt to stimulate a discussion that might lead to such a consensus, and to promote awareness of these issues, a group of astronomers recently established "An Astronomers' Data Manifesto".

In this paper, I discuss the successes and challenges of astronomical data management, and describe the manifesto and its purpose.

## 2 THE ASTRONOMICAL LITERATURE AND DATA CENTRES

### 2.1 The Astronomical Literature

Virtually all papers in the fields of astronomy, astrophysics, and related areas are referenced by the Smithsonian/NASA Astrophysics Data System (http://www.adsabs.harvard.edu/), known colloquially as the ADS. It references not only the mainstream journals, but conference reports, theses, preprint servers, and even institutional technical reports, where they are made public. It includes links to the papers in all their published forms, so that, for astronomers with institutional access to the journals, this provides transparent access to the entire astronomical literature. Even authors who publish papers (such as this) in non-astronomical journals can request to have their paper listed by the ADS. Powerful facilities enable a search to be made by author, title, keyword, or even text contained within the paper. As a result, the ADS has probably become the primary entry point to the published literature for most astronomers.

To disseminate their new research results, most authors now submit preprints (usually after acceptance by a journal) to arXiv.org (http://www.arxiv.org/list/astro-ph/new). This has become the primary means of accessing new research results, and many astronomers check it daily. Basic search facilities are also available, although ADS probably remains the most flexible way of accessing the arXiv.org contents.

The principal commercial astronomical journals have responded positively to these changes, and their electronic editions have become the main journals of record. It is likely that paper editions will be phased out within a few years. However, it is unclear how the astronomical publishing paradigm will change, given a number of conflicting forces:

1. There is a growing demand for open access, or free, journals, particularly as a solution to the "Digital Divide" discussed below. However, it is not yet clear how an open-access journal can afford to maintain the editorial quality and peer-review processes currently offered by mainstream journals.

2. Since most astronomers now access the literature via ADS or arXiv, the "title" of a journal has become less important. While there is still prestige associated with publishing in a high-impact journal, the actual visibility is similar regardless of where the paper is published, and so in time the impact factor may cease to differentiate journals. The commercial journals will therefore need to offer additional value, compared to open access journals, if they are to retain their authors and readers.

3. Some of the main journals have an excellent track record of responding to the changing demands of the astronomical community, and of promoting initiatives such as electronic access to associated data and tables, and linkages to other data centres. As a result, there is a significant groundswell of support for such journals from within the astronomical community.

## 2.2 Astronomical Data Centres

Astronomy enjoys a number of first-class data centres, the best known of which are CDS and NED.

NED is the NASA/IPAC Extragalactic Database (http://nedwww.ipac.caltech.edu/), which offers access to data taken from the literature and from major astronomical surveys. Its search engine provides all available data on an object or position in the sky, for which it will list measured data, images, references to the literature, and even some interpretation by comparing measurements made at different wavelengths by different authors and instruments (Fig 1). The difficulty of accomplishing this latter feat should not be underestimated, as authors use different metadata, jargon, and (even if they don't know the word) ontologies. As well as providing access to data, NED also provides a number of tools and innovative facilities such as its knowledgebase. Its key constraint is that it is designed to include only extragalactic objects (i.e. objects lying outside the Milky Way) and so does not include, for example, stars within the Milky Way, or solar-system objects. Nevertheless, for those who focus on extragalactic astronomy, NED has become the primary tool for accessing data.

CDS is the Centre de Données Astronomiques de Strasbourg (http://cdsweb.u-strasbg.fr/). Like NED, it offers access to data from the literature and from major surveys, and provides tools and search engines to access and interpret that data. It differs from NED in that it aims to include data on all astronomical objects outside the solar system, whether extragalactic or galactic. The main databases at CDS are Vizier, which includes nearly all major published surveys and tables, and Simbad, which provides search tools to access data taken both from the literature and form surveys. A number of other powerful tools are also provided, such as Aladin which enables a user to superimpose images from several data sources, including personal files. Just as important are the CDS research and development programs, which have been influential in shaping the way in which astronomers use data, and continue to be important drivers in the development of the VO.

A number of other major data centres around the world, such as those in Canada, China, Japan, and Russia, offer significant features for particular purposes. In addition, a number of specialised data centres exist to serve data for particular instruments or classes of instrument, such as NASA's High Energy Astrophysics Science Archive Research Center (http://heasarc.gsfc.nasa.gov/). Furthermore, the electronic data provided by the journals themselves effectively constitute a data centre, a blurring which is increasing as journals explore innovative projects such as those which offer to store authors' source data. Finally, it is important to acknowledge the regrettable closure of a major data centre (NASA's Astrophysical Data Center) in 2002, which serves as a warning against any complacency that high-performing data centres are immune to threats of closure.

## 2.3 Linkages between Literature and Data Centres

Many astronomers assume that data centres such as NED and CDS give them access to essentially all the published data. However, Andernach (2006), who has conducted a case study of over 2000 published articles, finds that typically only about 50% of results published in journals ever appear in the data centres, and lists some surprising and significant omissions.

It is not hard to understand why. At present, when authors submit data such as tables, spectra, or images, to journals, they do so in a variety of formats. The meaning of the axes or columns is often only apparent after reading the captions or the body of the paper, and authors continue to use jargon which is opaque to anyone outside the immediate field. Even worse, formatting errors still occur in published data tables, further impairing attempts at machine-readability.

To incorporate these data into a data centre requires a knowledgeable expert to interpret the words of the author, so that the results can be translated into a standard form. Given the finite resources of the data centres, and the expanding volume of astronomical literature, the availability of such knowledgeable experts then becomes a bandwidth bottleneck between the literature and the data centres.

Naturally, users would like to see all peer-reviewed results appear in the data centres. This will become increasingly important as VO tools become more widely used. In the same way that, at the moment, an astronomical publication that does not appear in ADS is effectively invisible to the astronomical community and will probably never be cited, in a few years time a result or image that is not accessible by the VO may never be used, generating wasteful repeated observations and slowing down the rate of scientific discovery.

How can we ensure that all validated astronomical data appear in the data centres? One solution would be to increase funding to the data centres so that they can employ enough knowledgeable experts to interpret all published data, but the finite available resources make this option unlikely.

An alternative is to find ways of automatically transferring published data from journals to data centres. This would probably require that the authors provide data in standard formats and that they provide the necessary metadata to interpret them. Then, if an author chooses to supply these metadata, and certifies that the data have been checked using appropriate tools, they could be imported automatically into the data centres.

This effectively redistributes the transcription workload from the data centres to the authors, and necessarily entails more work for authors. However, they benefit from the greater scientific impact and the higher citation rate that will result from their data being in the data centres. In many cases the paper itself will benefit from this further level of checking.

There is a potential disadvantage to such a system, in that it increases the likelihood that simple formatting errors in published papers might ultimately reduce the quality of the data in the centres. It remains to be seen whether automated checking procedures can reduce this possibility to the level where the disadvantage is outweighed by the advantage of doubling the quantity of high-quality information offered by the data centres.

## 3 OPEN ACCESS

Most astronomical data are unfettered by intellectual property or confidentiality issues, other than widely supported exceptions such as initial protection of observers' data by major facilities. As a result, astronomical archive data are generally available to all astronomers at no charge. It is this tradition which has enabled the success of astronomical data centres, and which will be vital for the success of the VO. The adoption of an open-access policy is not just for public good. For example, the Hubble archive results in roughly three times as many papers as those based on the original data (Beckwith, 2004). Similarly, the International Ultraviolet Explorer (IUE) archive increased the usage of IUE data by a factor of 5 (Wamsteker & Griffin, 1995). So, in principle, observatories might multiply their scientific output by making their archive data public. Since the funding for most major observatories depends on performance indicators such as publications and citations, it may be an expensive decision for an observatory not to adopt an open-access policy.

Because of such considerations, the 2003 General Assembly of the International Astronomical Union (IAU) adopted a resolution which says, broadly, that publicly-funded archive data should be made publicly available. This is closely aligned with recommendations by the International Council for Science (ICSU) and the Organisation for Economic Cooperation and Development (OECD, 2004), and may be regarded as a first step towards articulating the principles by which the astronomical community would like to see its data managed. Since then, a number of observatories, notably the European Southern Observatory, have embraced an open-access policy, but there remain a number of observatories that have not yet made their archival data publicly available, typically because of resource constraints. There also remain a few observatories (mainly privately-funded) which allow data archive access only to affiliated scientists, while still benefiting from the open access policies of other institutions.

Of course, astronomy is not alone is promoting Open Access. The ICSU's Committee on Data for Science and Technology (CODATA) has started the "Global Information Commons for Science

Initiative" whose aim is "open access and re-use of publicly-funded scientific data and information, and … cooperative sharing of research tools and materials among researchers" (CODATA, 2003).

## 4 THE VIRTUAL OBSERVATORY

The goal of the Virtual Observatory (VO) is to provide astronomers with seamless access to on-line resources, which might include data, tools, services, or theoretical models. The VO project actually consists of several national VO projects, linked together by the International Virtual Observatory Alliance (IVOA). Naturally, such a project relies heavily on standards for interoperability, and the definition of those is coordinated by the IVOA.

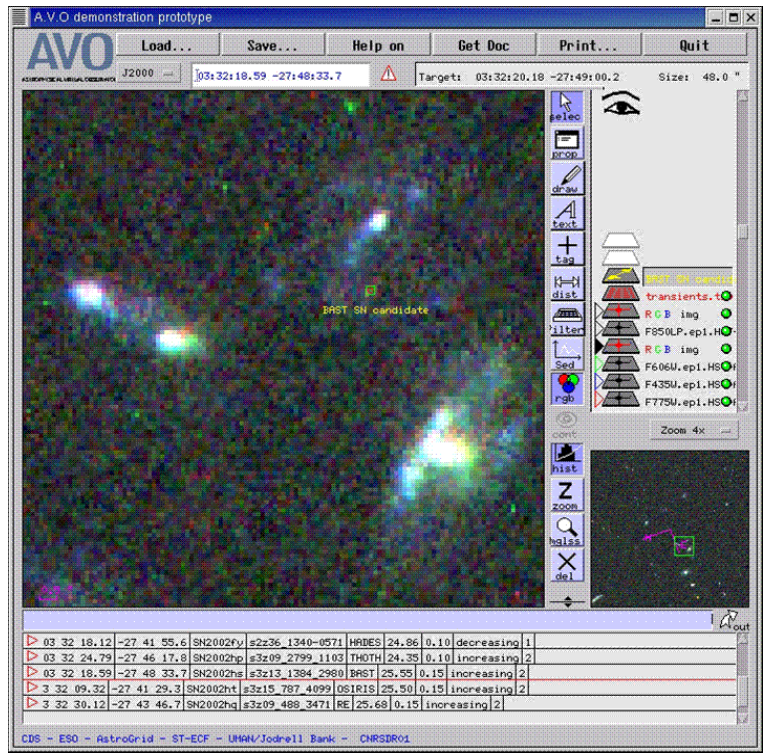

**Figure 2:** A screen-shot of a demonstration tool built for the Virtual Observatory, which aims to make all astronomical data available to all astronomers from their desktop, as easily as Google.

Since the IVOA was formed in 2001, it has enjoyed considerable success in defining those standards and coordinating the construction of tools and services. As a result, several demonstration applications are now available (Fig.2). A community of VO service providers is now being established, which will help data providers use the VO framework, and gather their feedback from implementation. Remaining challenges include long-term data access, data quality and curation certification, version control, histories, and intellectual property standards. More information about the VO can be found on http://www.ivoa.net/.

## 5 NEW FACILITIES

Much of the success of current astronomy is due to new instruments, ranging from the high-budget Space Telescopes such as Hubble, Chandra, and Spitzer, to low-budget university instruments which make important niche contributions. Whilst planning processes for space observatories naturally include provision for data processing, data management, and archiving, these important aspects are often neglected in smaller instruments, whose ambition sometimes exceeds their budget.

Experience with successful projects such as SDSS (Sloan Digital Sky Survey: http://www.sdss.org/) and 2MASS (Two Micron All Sky Survey: http://pegasus.phast.umass.edu/) suggests that roughly half the cost of a modern ground-based telescope is spent on the software, data processing, and data management. Smaller projects tend to neglect this, instead allocating the entire construction budget to telescope hardware, and expecting graduate students or support staff to figure out how to manage the data once they arrive. This results in instruments which perform well technically, but which fail to deliver the expected scientific impact. Experience suggests that the benefit/cost ratio for a new telescope is maximised by explicitly planning and budgeting for data management.

## 6 THE DIGITAL DIVIDE

The term "Digital Divide" refers to the gulf between those who have high-bandwidth access to information, data, and web services, and those, typically in developing countries, who do not. Without intervention, the gulf widens, because the lack of access results in further disadvantages, making high-bandwidth access in the future even less likely. The problem is not simply a humanitarian one. There is also the pragmatic problem that a large fraction of the world's potential intellectual resources are on the wrong side of the Digital Divide, and we would accelerate the rate of scientific discovery if we could tap into that resource.

Astronomers are better positioned to attack the digital divide than their colleagues in some other disciplines, because so much of the astronomical data and literature are freely or cheaply available on-line. Powerful VO tools and services will provide even better access, provided the internet connection is available. Meanwhile, facilities such as SALT and GMRT are already demonstrating the feasibility of building leading-edge facilities in developing countries.

A successful model is that of the Indian Inter-University Centre for Astronomy and Astrophysics (IUCAA). Associates and their students from all over India are funded to spend a few months every year at IUCAA, where they develop their own research programs and set up collaborations. The resulting technology and expertise are then transferred back to their home universities. The IUCAA represents an excellent model that might be replicated elsewhere in the developing world, and demonstrates that, although the digital divide extends throughout a society, astronomy is well-positioned to challenge it.

## 7 LEGACY DATA

The quality and quantity of digital data produced by modern observatories make it easy to neglect astronomy's rich legacy of photographic observations. Why should we bother saving these older data, which are usually so inferior to data from modern observatories?

The trouble is that astronomical objects change with time, and we cannot reproduce historical observations of an object which has disappeared or changed. For example, when the nearest supernova since the invention of the telescope exploded in 1987, astronomers were fortunate in being able to access old photographic plates which showed the host star before it disappeared in a supernova, as a result of which our knowledge of supernova processes is now enormously greater. Similarly, Captain Cook's eighteenth-century ship logs have provided invaluable data to help understand global warming, and historic stellar spectra have provided information on the Earth's ozone concentrations.

However, resources for safeguarding and digitising these data are necessarily in competition with those required to generate new data, and it is unlikely that funding will ever be available to digitise every photographic plate. Meanwhile the plates and their metadata are being lost or damaged. It is therefore important to determine, as far as we can, what value to place on the historical archives, and to determine a workable solution for their long-term storage and digitization before we lose the opportunity to make that decision.

The second area of concern is that we may be setting up a similar problem for future generations. The migration of present-day digital data is now well-managed in data centres, so that as technology evolves, data are migrated seamlessly to new media or new formats. Outside the data centres, however, the problem remains. Individual astronomers and small observatories keep magnetic tapes, such as DAT and Exabyte tapes, well beyond their recommended lifetime. Few now have the technology to read a round magnetic tape or a 5 1/4 -inch floppy. How long before an Exabyte, or even a CDROM or a DVD, becomes unreadable?

Along with other observational sciences, astronomy needs to take a broader view of how it safeguards its data.

# The Astronomers' Data Manifesto

**We, the global community of astronomy, aspire to the following guidelines for managing astronomical data, believing that they would maximise the rate and cost-effectiveness of scientific discovery.**

1. All significant tables, images, and spectra published in journals should appear in the astronomical data centres.
   - Access to data centres (CDS, NED, ADS, etc.) has significantly increased our scientific productivity. That productivity would be even higher if those data centres included all the data published in astronomical journals. That can not happen within our present system and existing resources.
   - Journals, data centres, and users should collaborate to define formats, table descriptions, and metadata that are easy for authors to adhere to, and can automatically be entered by the data centres into their databases. Authors are already required to adhere to strict formats for bibliographic references, but are not asked to adhere to standards for data tables.
   - Authors should be invited by the journals to submit electronic tables, spectra, and images to journals using these standard machine-readable formats together with metadata. If they choose to do so, their data will automatically be transferred to the data centres, resulting in greater effectiveness of their results, and presumably an increased number of citations. If they choose not to do so, then, as at present, there is no guarantee that their table will appear in a data centre.
   - In order that the required metadata are available to authors, it will also be necessary for observatories to ensure that all required metadata (e.g. filter wavelengths) are stored in data produced by the observatory. We note that this is already regarded as best practice by the world's leading astronomical observatories.

2. All data obtained with publicly-funded observatories should, after appropriate proprietary periods, be placed in the public domain.
   - We recognise the additional cost of doing so, but note that several major facilities have found this to be a cost-effective way of generating additional science per taxpayer dollar.
   - This carries an implicit requirement for appropriate user interfaces and data formats, which is within the brief of the Virtual Observatory and the data centres.
   - This principle was adopted as a Resolution by the IAU at the XXVth GA in Sydney, and is aligned with OECD and ICSU resolutions.

3. In any new major astronomical construction project, the data processing, storage, migration, and management requirements should be built in at an early stage of the project plan, and costed along with other parts of the project.
   - This may seem a statement of the obvious, especially to those major projects that already routinely follow this practice. But not all projects have done so, resulting in instruments which perform well technically, but which fail to deliver the expected level of science.

4. Astronomers in all countries should have the same access to astronomical data and information.
   - Major astronomical journals and data centres should provide free or reduced-cost electronic access to institutions in developing countries.
   - Where broadband internet access is not available to institutions, the IAU should work with other agencies and organisations to facilitate that access.

5. Legacy astronomical data can be valuable, and high-priority legacy data should be preserved and stored in digital form in the data centres.
   - Funding for these activities competes with funding for new instruments, and we recognise the need to demonstrate the value of the preserved data.
   - We need to reach a community consensus as to which data should be preserved, digitised, and migrated.
   - Time-variable phenomena, and of objects or events that cannot be re-observed, are amongst the highest priority, but we recognise the difficulty of establishing which data are likely to be most valuable in the future.

6. The IAU should work with other international organisations to achieve our common goals and learn from our colleagues in other fields.
   - Other fields of science are tackling similar issues, and some of our challenges are common to all areas of science. The IAU is the appropriate body to build and maintain the global networks and linkages necessary to attack the common problem.
   - An example of a common goal is to preserve the ability to place public domain scientific databases on the internet, which is deprecated by some groups concerned with the licensing of intellectual property.
   - ICSU and CODATA, with the participation of their member scientific unions, are actively involved in many of the issues discussed here, and the IAU should become a full participant in such activities.

We do not underestimate the challenge, but believe that these goals are achievable if astronomers, observatories, journals, data centres, and the Virtual Observatory Alliance work together to overcome the hurdles.

**Figure 3:** The Astronomers' Data Manifesto

## 8 CROSS-FERTILISATION WITH THE BROADER SCIENTIFIC COMMUNITY

There is a tendency in any scientific discipline to regard activity in other disciplines as irrelevant. After all, the data needs of an astronomer may seem very different from those of a geoscientist or a biologist. The underlying challenges, however, are remarkably similar, so a scientist in one discipline can learn much from those in other disciplines.

CODATA (the International Council for Science's Committee on Data for Science and Technology) provides one such forum where such cross-fertilisation may take place. CODATA's "Global Information Commons for Science Initiative" is particularly well-aligned with the changes taking place in astronomy, and astronomers would do well to join enthusiastically with the Global Information Commons movement.

Such cross-fertilisation need not take place only at an international level. Other communities such as geosciences and bioinformatics are discussing similar issues at national and local levels, and experience demonstrates that astronomers' participation in those discussions is fruitful for both sides.

## 9 THE DATA MANIFESTO

In an attempt to raise awareness of these issues, and define the goal, the IAU Working Group for Astronomical Data proposed the Astronomers' Data Manifesto, shown in Figure 3. This is intended not as a rigid declaration, but as a stimulus for discussion, which will hopefully converge to a consensus on how the astronomical community would like to see its data managed.

In writing this manifesto, we were aware of the many technical and sociological challenges implicit in the key points, and we do not underestimate the difficulty of achieving a solution. Nevertheless we believe that these goals are achievable if the astronomical community works together to overcome the hurdles.

## 10 CONCLUSION

Astronomy's success in recent years can be attributed partly to forward-thinking data initiatives, and the VO provides an opportunity for even greater success in the future. However, the success of the VO will rest on sound data management practices, and at present many astronomical facilities fall short of this.

One reason is that many astronomers and observatory directors simply don't know what good data management looks like, partly because there is no global consensus on what it should look like in astronomy. They are also unaware that good data management can generate good science, and that bad data management can inhibit the process of scientific discovery. While some major projects are shining examples of excellent data management, there are many examples where a lack of awareness of the issues is resulting in the loss or under-utilisation of data, leading to a decrease both in scientific cost-effectiveness and in the quality of the science. The problem is exacerbated by the fact that some astronomers find such issues unexciting compared to the thrill of the hunt for new discoveries.

To increase awareness and engagement, an "Astronomers Data Manifesto" has been formulated with the aim of initiating a discussion that will lead to the development of a code of best practice in astronomical data management. So far, this manifesto has been used successfully as a starting point for discussions. In time, some of these discussions will become solid proposals, leading to the adoption of resolutions or the initiation of technical projects. Eventually, we hope that the manifesto will become outdated, its propositions so obvious and widely accepted that it will be difficult to conceive of a time when such a manifesto was necessary. Meanwhile we still hear the occasional astronomer asking, in private, "Why should I share my data with my competitors?"

These issues faced by astronomy are common to many areas of science, and already the manifesto has received keen interest from biologists and geoscientists. Similar initiatives are developing in other fields, together with cross-disciplinary movements such as CODATA's Global Information Commons for Science Initiative. This cross-fertilisation promises to strengthen all fields of science - we have much to learn from each other.

## ACKNOWLEDGEMENTS

I thank my colleagues in the IAU Working Group on Astronomical Data, the participants in the "AstroData" e-discussions, and the participants in discussions at the IAU 2006 General Assembly, for their contributions which led to the ideas expressed in this paper. I particularly thank the co-authors of Norris et al. (2006), on which this paper draws heavily, and several of whom gave valuable comments on an earlier draft of this paper. However, the opinions expressed here are entirely my own.